\documentclass[twocolumn,aps,superscriptaddress]{revtex4-1}

\usepackage{palatino}
\usepackage[latin1]{inputenc}
\usepackage{epsf}
\usepackage{amsmath,amssymb}
\usepackage{latexsym}
\usepackage{calc}
\usepackage{color}
\usepackage{graphicx}
\usepackage{braket}

\usepackage[normalem]{ulem}

\newcommand{\ben}{\begin{equation}}
\newcommand{\een}{\end{equation}}
\newcommand{\bea}{\begin{eqnarray}}
\newcommand{\eea}{\end{eqnarray}}

\def\sss{\scriptscriptstyle\rm}

\def\1s{_{1,\sss S}}
\def\2s{_{2,\sss S}}

\begin{document}
\title{ Effect of Many Modes on Self-Polarization and Photochemical Suppression in Cavities}
\author{Norah M. Hoffmann}
\email{norah.magdalena.hoffmann@mpsd.mpg.de}
\affiliation{Department of Physics, Rutgers University at Newark, Newark, NJ 07102, USA}
\affiliation{Max Planck Institute for the Structure and Dynamics of Matter and Center for Free-Electron Laser Science and Department of Physics, Luruper Chaussee 149, 22761 Hamburg, Germany}
\author{Lionel Lacombe}
\email{liolacombe@gmail.com}
\affiliation{Department of Physics, Rutgers University at Newark, Newark, NJ 07102, USA}
\author{Angel Rubio}
\email{angel.rubio@mpsd.mpg.de}
\affiliation{Max Planck Institute for the Structure and Dynamics of Matter and Center for Free-Electron Laser Science and Department of Physics, Luruper Chaussee 149, 22761 Hamburg, Germany}
\affiliation{Center for Computational Quantum Physics, Flatiron Institute, 162 5th Avenue, New York, NY 10010, USA} 
\affiliation{Nano-Bio Spectroscopy Group and ETSF, Universidad del Pas Vasco,
20018 San Sebastian, Spain}
\author{Neepa T. Maitra}
\email{neepa.maitra@rutgers.edu}
\affiliation{Department of Physics, Rutgers University at Newark, Newark, NJ 07102, USA}

\date{\today}
\begin{abstract}
The standard description of cavity-modified molecular reactions typically involves a single (resonant) mode, while in reality the quantum cavity supports a range of photon modes. Here we demonstrate that as more photon modes  are accounted for, physico-chemical phenomena can dramatically change, as illustrated by the cavity-induced suppression of the important and ubiquitous process of proton-coupled electron-transfer. Using a multi-trajectory Ehrenfest treatment for the photon-modes, we find that self-polarization effects become essential, and we introduce the concept of self-polarization-modified Born-Oppenheimer surfaces as a new construct to analyze dynamics. As the number of cavity photon modes increases, the increasing deviation of these surfaces from the cavity-free Born-Oppenheimer surfaces, together with the interplay between photon emission and absorption inside the widening bands of these surfaces, leads to enhanced suppression. The present findings are general and will have implications for the description and control of cavity-driven physical processes of molecules, nanostructures and solids embedded in cavities. 
\end{abstract}

\maketitle

The interaction between photons and quantum systems is the foundation of a wide spectrum of phenomena, with applications in a range of fields. One rapidly-expanding domain is cavity-modified chemistry, by which we mean here nuclear dynamics concomitant with electron dynamics when coupled to confined quantized photon modes~\cite{ebbesen2016hybrid,George2016,hiura2018vacuum,RTFAR18}. The idea is to harness strong light-matter coupling to enhance or quench chemical reactions,  manipulate conical intersections, selectively break or form bonds, control energy, charge, spin, heat transfer, and reduce dissipation to the environment, for example.
This forefront has has been strongly driven by experiments \cite{riek2015,moskalenko2015,George2016,byrnes2014,kasprzak2006,Schmidt2016,Ebbesen1,Ebbesen2}, with theoretical investigations revealing complementary insights \cite{Feist2015,galego2015,FRAR15,GGF16,Schachenmayer2015,Cirio2016,Flick2017a,schafer2019modification,RMLCY19,KBM16,TPV16,SHCCV18,GT18,HFTG17,V18,FSRAR18,TSP18,flick2016exact,RFPATR14,RTFAR18,schafer2018ab}. However, apart from a handful of exceptions~\cite{SSF19,MTEF1,MTEF2,Flick2017,PSCFG18,FHDKBKR19,FWRAR19,AMSTK20} the simulations of cavity-modified chemistry largely involve coupling to only one (resonant) photon mode, and the vast majority uses simple model systems for the matter part. The modeling of realistic cavity set-ups requires coupling to multiple photon modes that are supported in the cavity even if they are not resonant with matter degrees of freedom, and further, the description should account for losses at the cavity boundaries. Some strategies have been put forward to treat quantized field modes in the presence of dispersive and absorbing materials~\cite{HB92,B13,B132,BYW08,SB08} and theories have been developed to treat many modes and many matter degrees of freedom~\cite{SSF19,MTEF2,PSCFG18,FHDKBKR19,RFPATR14,PFTAR15,FSRAR18,FWRAR19,Flick2017,FRAR15}. So far unexplored however is an explicit demonstration of how the cavity-modified electronic-nuclear dynamics  change as one increases the number of photon modes in the simulation.

Molecules coupled to multiple photon modes represent high-dimensional systems for which accurate and computationally efficient approximations beyond model systems are needed. To this end,  the Multi-Trajectory Ehrenfest (MTE) approach for light-matter interaction has been recently introduced~\cite{MTEF1,MTEF2}, and benchmarked for two- or three-level electronic systems in a cavity. Wigner-sampling the initial photonic state to properly account for the vacuum-fluctuations of the photonic field while using classical trajectories for its propagation, this method is able to capture quantum effects such as spontaneous-emission, bound photon states and second order photon-field correlations \cite{MTEF1,MTEF2}. In particular, as the trajectories are not coupled during their time-evolution the  algorithm is highly parallelizable.  Therefore, due to the simplicity, efficiency, and scalability, the MTE approach for photons emerges as an interesting alternative or extension to other multi-mode treatments~\cite{SSF19,MTEF2,PSCFG18,FHDKBKR19,BYW08,RFPATR14,PFTAR15,FSRAR18}. \footnote{This includes Quantum-Electrodynamical Density Functional Theory (QEDFT)~\cite{RFPATR14,PFTAR15,FSRAR18,RTFAR18}, which is an exact non-relativistic generalization of time-dependent density functional theory that dresses electronic states with photons and allows to retain the electronic properties of real materials in a computationally efficient way.}

\begin{figure}[t]
\includegraphics[width=1.0\columnwidth]{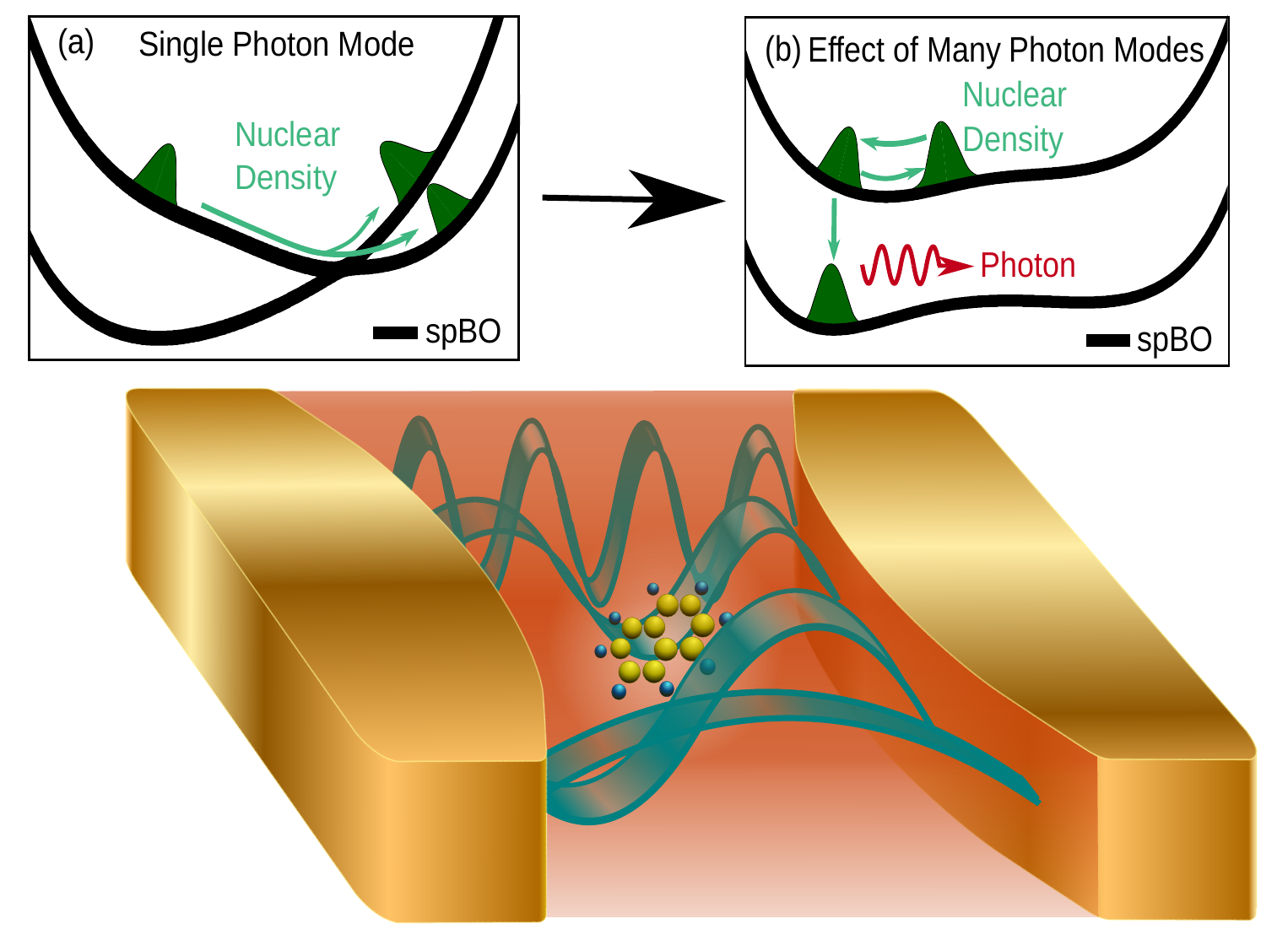}
\caption{An exemplary sketch of a molecule coupled to many photon modes. (a) Sketches the spBO surfaces and the corresponding nuclear dynamics for a coupling to a single photon mode. (b) Depicts the effect of many photon modes on the spBO surfaces and the corresponding complete photo-chemical suppression of the proton-coupled electron transfer.}
\label{fig:sketch}
\end{figure}

In this work, we extend the MTE approach to cavity-modified chemistry, and point out the effect that accounting for many photon modes has on coupled electron-ion dynamics. We focus on the process of polaritonic suppression of the proton-coupled electron transfer~\cite{OurPRL}, finding the electron-nuclear dynamics significantly depends on the number of modes, as sketched in Fig.~\ref{fig:sketch}. We neglect (for now) any effects from cavity losses so we can isolate effects purely from having  many modes in the cavity rather than a single mode.
To validate the MTE treatment of photons, we first study the single-mode case for which exact results can be computed, finding that MTE performs well but tends to underestimate the photon emission and cavity-induced effects. We explain why using the exact factorization approach~\cite{HARM18}. Treating also the nuclei classically gives reasonable averaged dipoles, and photon numbers, but a poor nuclear density, as expected. 
Turning to multi-mode dynamics computed from MTE, we find that as the number of cavity modes increases, the suppression of both proton transfer and electron transfer significantly increases (without changing coupling strength),  the electronic character becomes more mixed  throughout, and the photon number begins to increase. 
The results suggest that even when cavity modes are far from the molecular resonances, the chemical properties of the molecule can be dramatically altered by the presence of the cavity even when the coupling strength is not particularly large. 
The self-polarization term~\cite{SRRR19,rokaj2017,schafer2019modification} in the Hamiltonian that is often neglected in the literature, has an increasing impact on the dynamics, and we analyze the error made when neglecting it, depending on the number of photon modes accounted for. To this end, we introduce the concept of self-polarization-modified Born-Oppenheimer (spBO) surfaces as an instructive tool for analysis of chemical processes mediated by cavity-coupling.

\section{Hamiltonian}   
\label{sec:Hamiltonian}

In this work we consider the non-relativistic photon-matter Hamiltonian in the dipole approximation in the length gauge, i.e. applying the Power-Zienau-Woolley Gauge transformation \cite{PZ59} on the minimal coupling Hamiltonian in Coulomb gauge, as~\cite{T13,Flick2017a,RFPATR14,RTFAR18,HARM18,MKH20}
\begin{equation}
\hat{H}  = \hat{H}_m^{\rm SP} + \hat{H}_p + \hat{V}_{pm}\;,
\label{eq:fullH}
\end{equation}
with the Hamiltonian for the matter in the cavity as
\begin{eqnarray}
\hat{H}_m^{\rm SP} = \hat{T}_n + \hat{H}_{\rm BO}^{\rm SP}\; {\rm where}\;\;
\hat{H}_{\rm BO}^{\rm SP} = \hat{T}_e + \hat{V}_m + \hat{V}^{\rm SP}\;.
\label{eq:HspBO}
\end{eqnarray}
Our model is in one dimension,  with one electronic coordinate $r$ and one nuclear coordinate $R$, where the nuclear and electronic kinetic terms $\hat{T}_n = -\frac{1}{2M}\frac{\partial^2}{\partial R^2}$,  $\hat{T}_e = -\frac{1}{2}\frac{\partial^2}{\partial r^2}$, while $\hat{H}_{\rm BO}^{\rm SP}$ denotes the spBO Hamiltonian, defined by adding the  self-polarization term, 
\begin{equation}
\hat{V}^{\rm SP} = \frac{1}{2}\sum_\alpha^{\cal M} \lambda_\alpha^2(Z\hat{R} - \hat{r})^2\;,
\label{eq:Vsp}
\end{equation}
to the usual BO Hamiltonian. The self-polarization term depends only on matter-operators but scales with the sum over modes of the squares of the photon-matter coupling parameters $\lambda_\alpha$; thorough discussions of this term can be found in Ref.~\cite{SRRR19,rokaj2017,schafer2019modification}.
Atomic units, in which $\hbar = e^2 = m_e = 1$, are used here and throughout. The photon Hamiltonian and photon-matter coupling read as follows
\begin{eqnarray}
\hat{H}_p(q)&=& \frac{1}{2}\sum_{\alpha}^\mathcal{M}\left(\hat{p}_\alpha^2 + \omega_\alpha^2\hat{q}_\alpha^2\right)\\ \hat{V}_{pm}&=& \sum_{\alpha}^\mathcal{M}\omega_\alpha\lambda_\alpha\hat{q}_\alpha\left({Z\hat R} - \hat{r}\right) \;,
\end{eqnarray}
where $\alpha$ labels the photon mode, $\hat{q}_\alpha = \sqrt{\frac{1}{2\omega_\alpha}}(\hat{a}^\dagger_\alpha + \hat{a}_\alpha)$ is the photonic coordinate, related to the electric field operator, while $\hat{p}_\alpha = -i\sqrt{\frac{\omega_\alpha}{2}}(\hat{a} - \hat{a}^\dagger)$ is proportional to the magnetic field. It is important to note that in this gauge, the photon number is given by 
\begin{equation}
\hat{N}_p = \sum_\alpha \big(\hat{a}^\dagger_\alpha\hat{a}_\alpha + \lambda_\alpha \hat{q}_\alpha(Z\hat{R} - \hat{r}) + \lambda_\alpha^2(Z\hat{R} - \hat{r})^2/(2\omega_\alpha)\big)
\label{eq:Np}
    \end{equation}
    
We choose the matter-photon coupling strength through the 1D mode function
$\lambda_\alpha= \sqrt{\frac{2}{\cal{L} \mathrm{\epsilon_{0}}}}\sin(k_{\alpha}X)$ 
where $\cal{L}$ denotes the length of the cavity and $k_\alpha = \alpha\pi /{\cal L}$ the wave vector ( $\alpha = 1,2,3...$) , and $X$ the position measured from the center of the cavity. Unless stated otherwise, we take $X ={\cal L}/2$, assuming that the molecule is placed at the center of the cavity, and ${\cal{L}} = 12.5\mu m$,
much longer than the spatial range of the molecular dynamics. 
This cavity-length yields a coupling strength of $\lambda_{\alpha}=(-1)^{\frac{\alpha-1}{2}}0.01$a.u. for modes with odd $\alpha$, $\lambda_\alpha = 0$ for even $\alpha$, and a fundamental cavity mode of frequency $\omega_0 = 0.0018$ a.u.,
 and these parameters are used throughout the paper except for in benchmarking the single-mode results in Sec.~\ref{sec:MTE}. 

In our particular model the matter potential $\hat{V}_m$ is given by the 1D Shin-Metiu model \cite{SM95,FH97,FH97b}, which consists of a single electron and proton ($Z = 1$ above), which can move between two fixed ions separated by a distance $L$ in one-dimension. This model has been studied extensively for both adiabatic and nonadiabatic effects in cavity-free~\cite{FH97,FH97b,AASG13,AASMMG15} and in-cavity cases~\cite{OurPRL,GCGFF19,Flick2017a}. The Shin-Metiu potential is:
\begin{equation}\hat{V}_{m} = \sum_{\sigma = \pm 1}\left(\frac{1}{\vert R +\frac{\sigma L}{2}\vert} - \frac{{\rm erf}\left(\frac{\vert r + \frac{\sigma L}{2}}{a_\sigma}\right)}{\vert r + \frac{\sigma L}{2}\vert}\right) - \frac{{\rm erf}\left(\frac{\vert R - r\vert}{a_f}\right)}{\vert R - r\vert}
\end{equation}
We choose here  $L = 19.0$~a.u.,  $a_+ = 3.1$~a.u., $a_- =4.0$~a.u., $a_f = 5.0$~a.u., and the proton mass $M = 1836$~a.u.; with these parameters, the phenomenon of proton-coupled electron transfer occurs after electronic excitation out of the ground-state of a model molecular dimer~\cite{OurPRL}. Furthermore, for computational convenience in the MTE calculations we  truncate the electronic Hilbert space to the lowest two BO-surfaces.

\section{Self-Polarization-Modified BO Surfaces}
\label{sec:spBO}

Potential energy surfaces play a paramount role in analyzing coupled dynamics: we have Born-Oppenheimer (BO) surfaces for cavity-free dynamics, Floquet~\cite{Sambe73,FHS16} or quasistatic~\cite{TIW96,SKKF03} surfaces for molecules in strong fields, cavity-BO~\cite{Flick2017a} or polaritonic surfaces~\cite{galego2015} for molecules in cavities and the exact-factorization based time-dependent potential energy surface~\cite{AMG10,AMG12,OurPRL} for all cases that yields a complete single-surface picture.  The surfaces so far explored for molecules in cavities have largely neglected the self-polarization term, which is often indeed negligible for typical single-mode calculations.
However, its importance in obtaining a consistent ground-state and maintaining gauge-invariance has been emphasized~\cite{SRRR19,rokaj2017}. The self-polarization term involves a sum over the number of photonic modes considered \cite{de2018cavity}. In the multi-mode case, this sum can become as important as the other terms in the Hamiltonian, and, as we shall see below, it cannot be neglected, especially becoming relevant for large mode-numbers, contributing forces on the nuclei while the total dipole evolves in time. To analyze the dynamics, we define self-polarization-modified Born-Oppenheimer (spBO) surfaces $\epsilon_{\rm BO}^{\rm SP}(R)$, as eigenvalues of the spBO Hamiltonian: $\hat{H}_{\rm BO}^{\rm SP}\Phi_{R,\rm BO}^{\rm SP} = \epsilon_{\rm BO}^{\rm SP}(R)\Phi_{R,\rm BO}^{\rm SP}$. 

Further, we define "$n$-photon-spBO surfaces" by simply shifting the  spBO surfaces uniformly by  $n\hbar\omega_\alpha$. We note that this nomenclature should not be taken too literally since in the length gauge, the photon number includes matter-coupling and self-polarization terms on top of the Coulomb-gauge definition of $\langle a^\dagger a\rangle$ as shown in Eq.~(\ref{eq:Np}). That is, a $1$-photon-spBO surface does not actually denote a surface where there is one photon in the system, for example. 
The spBO surfaces can be viewed as approximate (self-polarization modified) polaritonic surfaces, becoming identical to them in the limit of zero coupling. For small non-zero coupling the polaritonic surfaces, defined as eigenvalues of $\hat{H} - \hat{T}_n$, resemble the $n$-photon-spBO surfaces when they are well-separated from each other, but when they become close, the crossings become avoided crossings.

The top middle panel of Fig.~\ref{fig:One-mode} shows the ($0$-photon) spBO (pink)  and $1$-photon spBO surfaces (black) for the single mode case, where the spBO and BO surfaces essentially coincide. The top panel of Fig.~\ref{fig:MultiPhoton1} shows in black the spBO surfaces for our system for 10, 30, 50, and 70 modes. For 10 modes, they show only a small deviation from the BO surfaces with a small widening and shift of the avoided crossing region. As the number of cavity modes grows, the spBO surfaces clearly show an increasing departure from the BO surfaces.
Given that the landscape of such surfaces provides valuable intuition about the nuclear wavepacket dynamics, with their gradients supplying forces, this suggests an important role of the self-polarization term in the dynamics of the nuclear wavepacket, as we will see shortly. 

The band-like structures indicated by the shaded colors in the top row of Fig.~\ref{fig:MultiPhoton1} represent the $1$-photon-spBO surfaces, forming a quasi-continuum. The shading actually represents parallel surfaces separated by the mode-spacing $0.0018$~a.u. (We note that, as a function of cavity-length, the mode-spacing decreases, approaching the continuum limit as $\cal{L}$ approaches infinity, however the coupling strength $\lambda_\alpha$ also decreases, vanishing in the infinite-$\cal{L}$ limit such that the free BO surfaces are recovered).
The $1$-photon ground-spBO band and $1$-photon excited-spBO band show growing width and increasing overlap as the number of photon modes increases, suggesting a nuclear wavepacket will encounter an increasing number of avoided crossings between ground- and excited- polaritonic states as it evolves.
It is worth noting that the $1$-photon-spBO band overlaps with the $(n>1)$-photon-spBO band of the lower frequencies, e.g. the 10-photon-spBO curve for the fundamental mode coincides with the 1-photon-spBO curve for the 10th mode. For simplicity however we will still refer to these as simply $1$-photon-spBO bands with the understanding that they may include some  higher-photon-number states for low frequencies. For clarity of the figure, we show only the $1$-photon-spBO band, but we note that the $(n>1)$-photon bands also play a role in the dynamics, in particular when there is overlap between the $(n>1)$-photon spBO ground state and the spBO excited state. 
We return to the implications of the spBO bands later in the discussion of the multi-mode cases. 

Finally, as mentioned above, the spBO surfaces do not incorporate the bilinear light-matter interactions, which if included would turn crossings of the spBO surfaces into avoided crossings of self-polarization modified polaritonic surfaces. Computing these for a large number of photon modes results in a large diagonalization problem. Instead, the spBO surfaces depend on only the matter operators and light-matter coupling strength so could in principle be computed with a similar computational expense as for BO surfaces while giving already an indication of how chemistry is modified in the cavity, as we will see shortly.

\section{MTE Treatment of Photonic System}
\label{sec:MTE}
A computationally feasible treatment of coupled electron-ion-photon dynamics in a multi-mode cavity calls for approximations. Here we will consider one electronic and one nuclear degree of freedom but up to 70 photon modes, so we use MTE for the photons, coupled to the molecule treated quantum mechanically.
 
We launch an initial Gaussian nuclear wavepacket on the excited BO surface at $R = -4$a.u. We take the initial state as a simple factorized product of the  photonic vacuum state $\xi_0(q)$ for each mode, the excited BO state, and the nuclear Gaussian wavepacket: $\Psi(r,R,{\underline{q}},0) = {\cal N}e^{-[2.85(R+4)^2]}\Phi^{\rm BO}_{R,2}(r)\xi_0({\underline{q}})$, where ${\underline{q}}$ denotes the vector of photonic displacement-field coordinates. More precisely, for the MTE for photons we sample the initial photonic vacuum state from the Wigner distribution given by: $ \xi_{0}(q,p)=\prod_{\alpha} \frac{1}{\pi}e^{\left[-\frac{p^{2}_{\alpha}}{\omega_{\alpha}} - \omega_{\alpha}q_{\alpha}^2\right]}$. Furthermore, with two electronic surfaces, the equations of motion are as follows, for the $l$th trajectory: 
\begin{align}
    \ddot{q}_{\alpha}^{~l}(t) = -\omega_{\alpha}^2q_{\alpha}^{~l} - \omega_{\alpha}\lambda_{\alpha}(Z\langle R \rangle^{l} - \langle r \rangle^{l}),\\
    i\partial_t\begin{pmatrix}     
   C_{1}(R,t)\\
    C_{2}(R,t) \end{pmatrix} = 
    \begin{pmatrix} 
    h_{11} & h_{12} \\
    h_{21} & h_{22}
    \end{pmatrix}
    \begin{pmatrix}
    C_{1}(R,t)\\
    C_{2}(R,t) 
    \end{pmatrix},
\end{align}
with the diagonal matrix elements 
\begin{align}
\nonumber
h_{ii} &= \epsilon_{\rm BO}^{i}(R) - \frac{1}{2M}\partial^{2}_{R} +\sum_{\alpha}\Big(\lambda_{\alpha}\omega_{\alpha}q_{\alpha}^{l}(ZR-r_{ii}(R)) \\ &+ \frac{\lambda_\alpha^2}{2}\cdot((ZR)^2 - 2ZRr_{ii}(R) + r^{(2)}_{ii}) \Big)
\end{align} and for $i\neq j$,
\begin{align}
\nonumber
h_{ij} &= -\frac{1}{M}d_{ij}(R)\partial_{R}- \frac{d_{ij}^{(2)}(R)}{2M} -\\ &\sum_{\alpha}\lambda_{\alpha}\omega_{\alpha}q_{\alpha}^{l}r_{ij}(R) + \sum_{\alpha}\left(\frac{\lambda_{\alpha}^{2}}{2}\cdot\left(-2ZRr_{ij}(R) + r_{ij}^{(2)}(R)\right)\right)
\end{align}
 Here the non-adiabatic coupling terms are 
 $d_{ij}(R) = \langle \Phi_{R,i}^{\rm BO} \vert \partial_{R} \Phi_{R,j}^{\rm BO} \rangle$, $d^{(2)}_{ij}(R) = \langle \Phi_{R,i}^{\rm BO}  \vert \partial^{2}_{R} \Phi_{R,j}^{\rm BO} \rangle$, and the transition dipole and quadrupole terms $r_{ij}^{(n)} = \langle \Phi_{R,i}^{\rm BO} \vert \hat{r}^n\vert\Phi_{R,j}^{\rm BO} \rangle$.
 The coefficients $C_i(R,t)$ are the expansion coefficients of the electron-nuclear wavefunction in the BO basis: $\Psi(r,R,t) = \sum_{i = 1,2}C_i(R,t)\Phi^{\rm BO}_{R,i}(r)$. Subsequently, $R$-resolved and $R$-averaged BO-populations are defined as $\vert c_{1,2} (R,t)\vert^2 = \vert C_{1,2}(R,t)\vert^2/\vert\chi(R,t)\vert^2$ and $\vert C_{1,2}(t)\vert^2 = \int dR \vert C_{1,2}(R,t)\vert^2$ respectively. In the single-mode case we will  also present the results for when the proton is also treated by MTE with the nuclear trajectory satisfying $M\ddot{R}^{l}(t) = -\langle\partial_{R}\epsilon_{\rm BO}(R^{l})\rangle - \sum_{\alpha}\omega_{\alpha}\lambda_{\alpha}q_{\alpha}^{l} - \sum_{\alpha}\left(\lambda_{\alpha}^2Z(Z\langle R\rangle^{l} - \langle r\rangle^{l})\right) $. 
 For all MTE calculations 20,000 trajectories were enough for convergence the results for all cases.

\section*{Results}

\subsection{Single-Mode Benchmark}
\begin{figure}
\includegraphics[width=1.0\columnwidth]{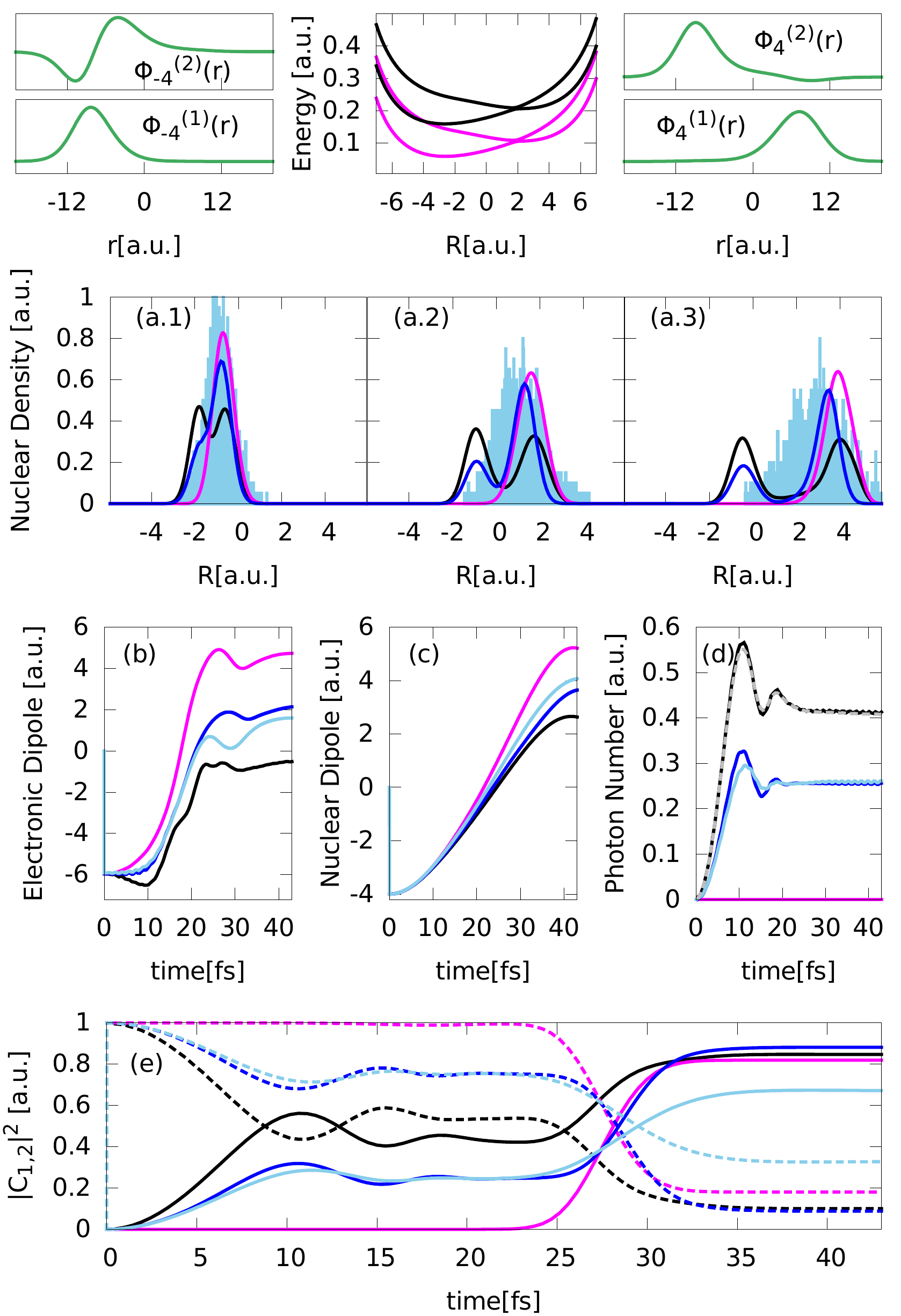}
\caption{Single-Mode Case: The top panel shows the ground (lower) and excited (upper) BO wavefunctions at $R=-4$~a.u. (left) and at $R=4$~a.u. (right) and the spBO surfaces (pink) and one-photon spBO surfaces (black). The spBO surfaces are essentially identical with the BO surfaces however for the single-mode case. The second panel depicts the nuclear density for cavity-free (pink), full quantum treatment (black), MTE treatment of the photons only (blue) and MTE treatment of both photons and nuclei (light blue) at time snapshots $t=22$~fs (a.1) ,$t=30$~fs (a.2) and $t=38$~fs (a.3). The third panel shows the electronic (b) and nuclear (c) dipole and the photon number (d). In panel (d) the dashed-grey line shows the photon number Eq.~(\ref{eq:Np}) while the black shows the first term only.  The lowest panel depicts the BO occupations, $\vert C_{1,2}(t)\vert^2$.}
\label{fig:One-mode}
\end{figure}

First we consider a single-mode case for which we are able to compare the MTE method to numerically exact results \footnote{We note that the two-mode case can also be solved exactly numerically, but the single-mode comparison here already illustrates the main points.}. The cavity-free dynamics of our system shows "proton-coupled electron transfer" in the following sense: The top panel of Fig.~\ref{fig:One-mode} shows the electronic wavefunctions at $R=-4$~a.u. (left) and $R=4$~a.u. (right) in the cavity-free case, showing that the transition of the initial nuclear wavepacket  to the lower BO surface through non-adiabatic coupling near the avoided crossing results in an electron transfer. 
Ref.~\cite{OurPRL} found that this proton-coupled electron transfer is suppressed when the molecule is placed in a single-mode cavity resonant with the initial energy difference between the BO surfaces. This energy difference is $0.1$~a.u. which would correspond to about the 56th mode in the cavity with the parameters described in Sec.~\ref{sec:Hamiltonian}. A single mode of frequency equal to the fundamental mode of that cavity is so far off the initial resonance that the dynamics is only slightly altered from the cavity-free case (see Sec.~\ref{sec:MTE}).
Instead, for the purposes of benchmarking the MTE method for cavity-modified dynamics in this section, we use the same parameters as in Ref~\cite{OurPRL}: cavity frequency of $0.1$~a.u. with light-matter coupling strength $\lambda = 0.005$~a.u.

The second row of Fig.~\ref{fig:One-mode} shows the dynamics of the nuclear wavepacket (see also supplementary materials, movie 1) for the exact cavity-free case (pink),   exact single-mode case (black), MTE for photons (blue) and MTE for both photons and nuclei (light blue). 
As discussed in Ref.~\cite{OurPRL}, the exact dynamics in the cavity shows suppression of proton-coupled electron transfer (compare pink and black dipoles in third panel), due to photon emission at early times (black line in panel (d)) yielding a partially trapped nuclear wavepacket, leading to less density propagating to the avoided crossing to make the transition to the lower BO surface. The BO-populations in the lowest panel (e) show the initial partial drop to the ground-state surface associated with the photon emission.

Both MTE approaches are able to approximately capture the cavity-induced suppression of the proton-coupled electron transfer, as indicated by the blue and light-blue dipoles and photon-number in panels (b--d), and approximate the BO occupations in panel (e) reasonably well. However 
 both approaches somewhat underestimate the suppression; the photon emission is underestimated by about a third, as is the suppression of the electronic dipole transfer,  for example. We note that the photon number is by far dominated by the first term in Eq.~(\ref{eq:Np}) (compare black and dashed gray line in panel (d)); there is only a single mode at an initially resonant molecular frequency, and the coupling is small enough that the second and third terms are very small.
  To understand why MTE underestimates the photon number, we compare the potentials the MTE photons experience to the exact potential acting on the photons as defined by the exact factorization approach, which was presented in Ref.~\cite{HARM18}. In this approach, the total wavefunction of a system of coupled subsystems is factorized into a single product of a marginal factor and a conditional factor, and the equation for the marginal satisfies a Schr\"odinger equation with potentials that exactly contain the coupling effects to the other subsystem. When the photonic system is chosen as the marginal, one obtains then the exact potential driving the photons, and this was found for the case of an excited two-level system in a single resonant mode cavity in 
 Ref.~\cite{HARM18}. It was shown that the potential develops a barrier for small $q$-values while bending away from an upper harmonic surface to a lower one at large $q$, creating a wider and anharmonic well. This leads then to a photonic displacement-field density with a wider profile in $q$ than would be obtained via the uniform average of harmonic potentials that underlie the MTE dynamics,  i.e. MTE gives lower probabilities for larger electric-field values, hence a smaller photon-number and less suppression compared to the exact.

An additional treatment of the nuclei within MTE yields a spreading of the nuclear wave packet instead of a real splitting (Fig.~\ref{fig:One-mode}(a.3)), a well-known problem of Ehrenfest-nuclei. This error is less evident in averaged quantities such as dipoles and BO coefficients. We note that an exact treatment of the photons coupled to MTE for only nuclei will not improve this situation. 
With more photon modes, the polaritonic landscape has even more avoided crossings, which are likely to make the Ehrenfest description for nuclei worse, calling for the development of more advanced propagation schemes~\cite{HLM18,MAG15}.

Having now understood the limitations of MTE, we now apply the MTE framework for photons to the multi-mode case.

\subsection{MTE Dynamics for Multi-Mode Cases}
\label{sec:multimode}
We return to the cavity-parameters of Sec.\ref{sec:Hamiltonian}, where the fundamental mode has $\omega_0 = 0.0018$ a.u. and the light-matter coupling strength $\lambda_\alpha = \pm 0.01$~a.u. for modes which are non-zero at the center of the cavity (see Sec.~\ref{sec:Hamiltonian}). We consider the effect on the dynamics as an increasing number of harmonics of the fundamental are included in the simulation: from 1, 10, 30, 50, to 70 modes. Although in principle all modes should be considered, already these cases demonstrate a dramatic impact of the number of modes on the dynamics. 

We note if we had instead used a cavity whose fundamental mode is $0.1$~a.u. as in the single-mode demonstration of the previous section, then considering the effect of including higher cavity-modes on the dynamics would be more complicated since one rapidly encounters cavity wavelengths short enough that the long-wavelength approximation is broken. Instead, with the parameters of Sec.~\ref{sec:Hamiltonian} that we use here for the many-mode cases, the maximum frequency included is $\omega_{max}=0.127[a.u.]$ in the $70$ mode case, which corresponds to a wavelength of $\lambda_{max}=0.057[\mu m]$, much larger than the spatial range of the molecular dynamics.

Another important aspect when including many photon modes is the well-known zero-point energy leakage problem. However, in all cases considered here we find none (for the 10 - 50 mode cases) or extremely small leakage for long times and high frequencies (for the 70 mode case) compared to the overall emission and photon number, with no impact on the dynamics. Still, as more modes are included (beyond 70 modes) we anticipate the zero-point energy leakage could become a problem and would need to be addressed carefully.

\begin{figure}
\includegraphics[width=1.0\columnwidth]{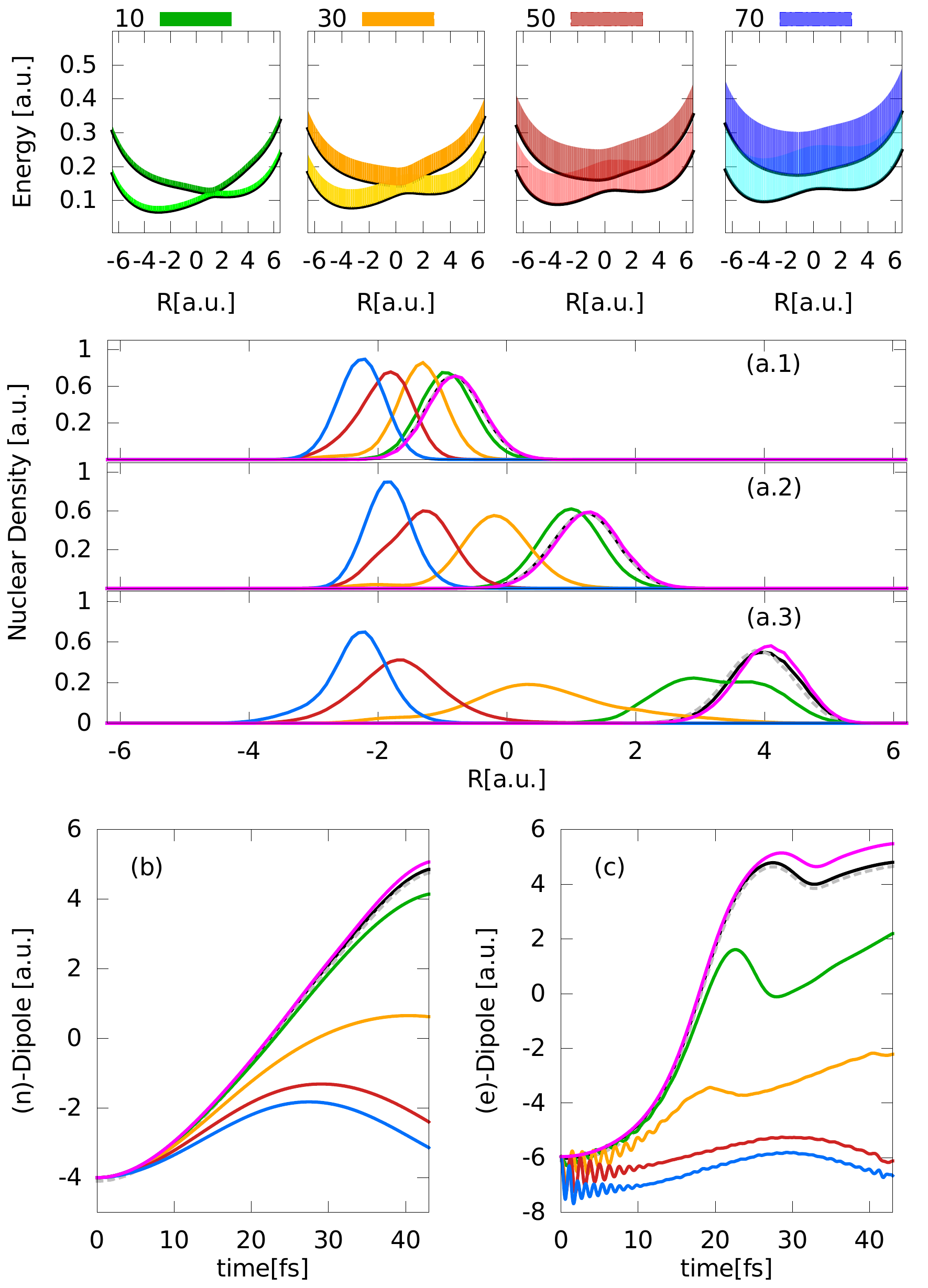}
\caption{The ground- and excited  $1$-photon spBO bands, representing surfaces separated by 0.0018 a.u. (see text) for 10 modes (green), 30 modes (orange), 50 modes (red) and 70 modes (blue). The middle panel depicts the nuclear density at time snap shots $t=22$fs (a.1), $t=30$fs (a.2) and $t=41$fs (a.3) in the same color code along with the single mode case computed within MTE-for-photons (black), exact (gray dashed) and cavity-free (pink) . The lowest panel shows the nuclear dipole (b) and electric dipole (c).}
\label{fig:MultiPhoton1}
\end{figure}

The top panel of Fig.~\ref{fig:MultiPhoton1} shows the ground and excited $1$-photon spBO bands, as introduced earlier. As mentioned in Sec.~\ref{sec:spBO}, we do not show the entire $(n>1)$-photon spBO bands explicitly for clarity, but it is important to note that the $1$-photon band does include some $(n>1)$-photon states of the lower frequency photon modes that are included in each simulation. 

As we observed earlier, including more photon modes has two effects on the spBO surfaces. First, the self-polarization morphs them away from the cavity-free BO surfaces, increasing their separation, and what was a narrow avoided crossing in the cavity-free case shifts leftward in $R$ with increased separation. Second,  the $1$-photon ground and excited spBO bands both broaden with increasing number of crossings with the $0$-photon spBO surfaces and with each other in the regions of overlap. As the gradient of these surfaces and the couplings between them are considerably altered, we  expect significant differences in the nuclear dynamics when going from the single-mode case to the many-mode case.

\begin{figure}
\includegraphics[width=1.0\columnwidth]{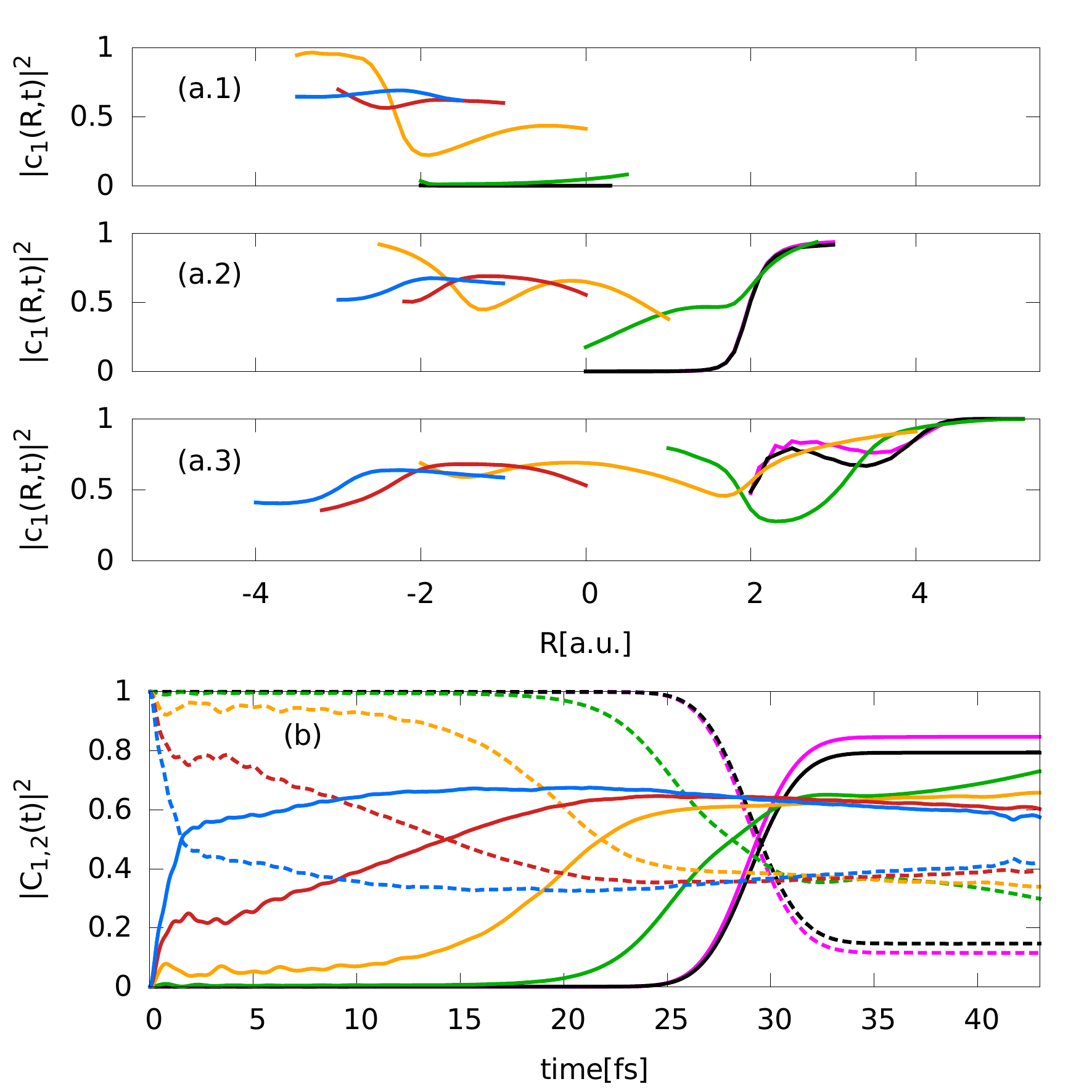}
\caption{Groundstate BO-surface population (a)  at time snap shots $t=22$~fs (1), $t=30$~fs (2) and $t=41$~fs (3) over $R$ and the averaged population over time (c) in the same color code as Fig.\ref{fig:MultiPhoton2}.}
\label{fig:MultiPhoton2}
\end{figure}

Indeed, this is reflected in the middle panel of Fig.~\ref{fig:MultiPhoton1} which shows the nuclear wavepacket  at time snapshots $22$~fs (a.1), $30$~fs (a.2),  $41$~fs (a.3) and in the lower panel, showing the nuclear dipole (panel b) and electronic dipole (panel (c)). The corresponding $R$-resolved BO-occupations of the ground-BO electronic state $\vert c_{1}(R,t)\vert^2$,  shown in Fig.\ref{fig:MultiPhoton2}(a), and the $R$-averaged occupations $\vert C_{1,2}(t)\vert^2$ over time plotted in Fig.\ref{fig:MultiPhoton2}(b) also show significant mode-number dependence (A movie is also provided in supplementary materials, movie 2).

\begin{figure}
\includegraphics[width=1.0\columnwidth]{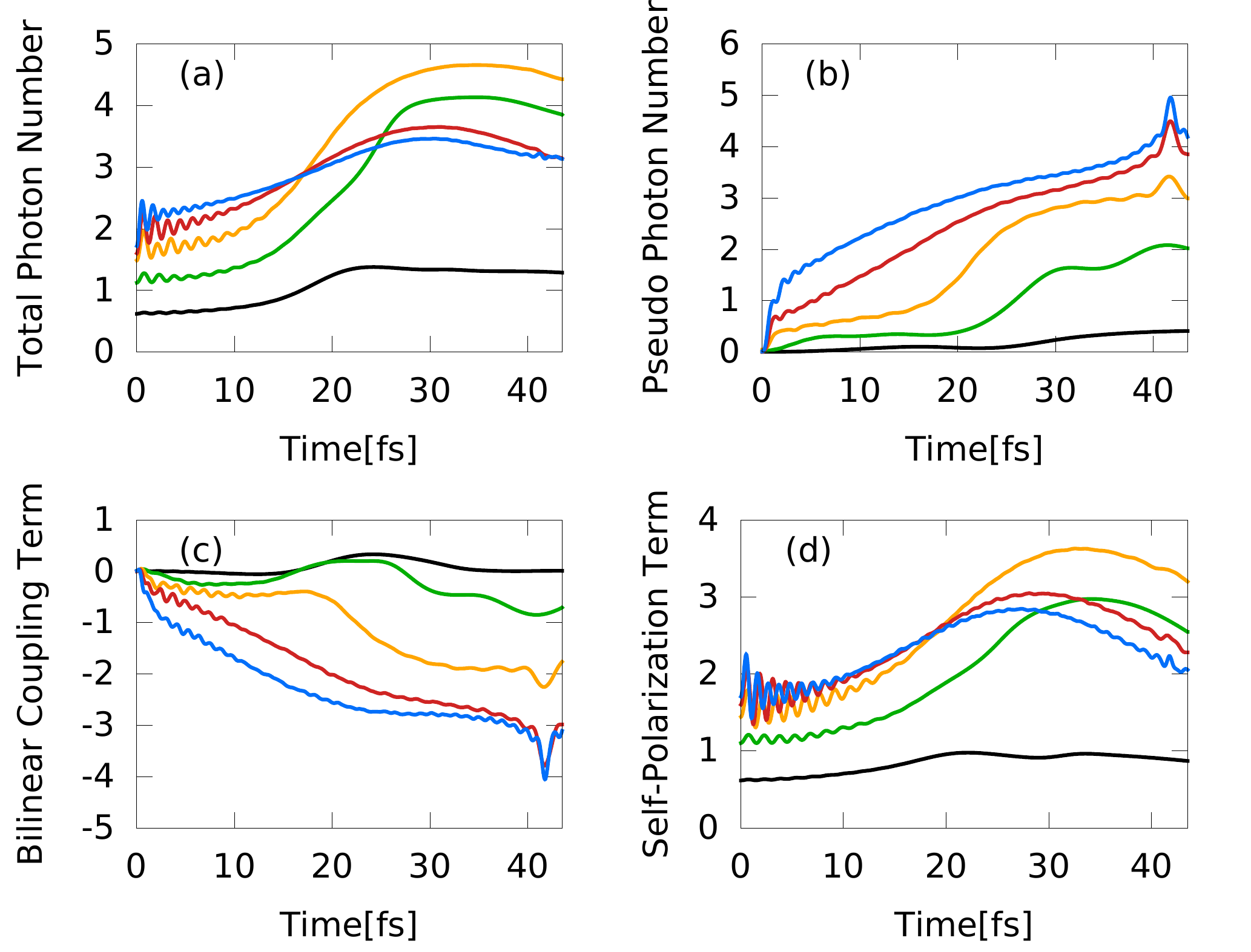}
\caption{The different components of the photon number: (a) Total photon number of Eq.~(\ref{eq:Np}) (b) Pseudo-photon number (first term in Eq.~(\ref{eq:Np})) (c) Bilinear coupling term (second term in Eq.~(\ref{eq:Np})), (d) Self-polarization term (third term in Eq.~(\ref{eq:Np})), using the same color code as Fig.\ref{fig:MultiPhoton2}.}
\label{fig:MultiPhoton3}
\end{figure}

Dynamics in the single-mode cavity (black as computed with MTE and gray dashed for exact, in Fig.~\ref{fig:MultiPhoton1}) is almost identical to the cavity-free case (pink), since the mode is far off the molecular resonance ($\omega = 0.0018$~a.u.) for the duration of the dynamics. Differences are seen when the nuclear wavepacket encounters the avoided crossing region, with the single-mode case slightly lagging behind the cavity-free dynamics, and with a smaller transfer to the lower electronic state. First, due to the stronger coupling ($\lambda = 0.01$~a.u.), compared to the single mode benchmarking, the spBO-surfaces (not shown here) already have a very slight distortion from its original BO-form, with a slightly wider and broader avoided crossing region. As a result, the transfer to the ground electronic state is slightly reduced, as evident in Fig.~\ref{fig:MultiPhoton2}(a.3, b) and Fig.~\ref{fig:MultiPhoton1}c). With more population in the upper state which slopes to the left after the avoided crossing, the wavepacket slows down compared to the cavity-free case (Fig.~\ref{fig:MultiPhoton1}a.3,b). 
The $0$-photon spBO surfaces at closest approach have an energy difference of about $0.006$~a.u. so the $1$-photon ground-state surface does not interact strongly with the excited spBO surface. The overlap of the $4$- and higher-photon ground-state surfaces with the excited spBO surface leads to a small photon emission as seen in Fig.~\ref{fig:MultiPhoton3}. We observe that unlike for the parameters of the previous section, the larger self-polarization term results in a significant difference between the true photon number of Eq.~(\ref{eq:Np}) and the ``pseudo-photon-number", the first term, even for a single-mode, but due to the low frequency of this mode there is only a limited impact on the energetics of the matter system. 

Going now to the 10-mode case (green in Fig.~\ref{fig:MultiPhoton1}), the spBO surfaces are visibly distorted from the BO-surfaces shown in Fig.\ref{fig:One-mode}, and we begin to see suppression of both the proton transfer in panels (a) and (b), and more so the electron transfer in panel (c). The largest cavity-frequency has a frequency $\omega_{max}=0.018$~a.u., while at closest approach the spBO surfaces differ in energy by $0.01$a.u. with their avoided crossing shifting further left to $R = 1.2$~a.u. The suppression of the molecular dynamics
begins to occur a little before the wavepacket approaches the avoided crossing between the spBO-surfaces, and is due to two effects: first, the gentler slope and weaker crossing of the spBO surfaces causing a weaker effective electron-nuclear non-adiabaticity, and second, the crossings between the $1$-photon ground spBO surfaces with the first excited spBO surface causing photon emission. These  crossings become avoided crossings once the matter-photon bilinear coupling is accounted for, i.e. in the polaritonic surfaces. At around $R=0.6$~a.u., a single photon of $\omega_{max}$ first becomes resonant to the self-polarization modified molecular excitation, enabling transitions from the 0-photon excited surface to the $1$-photon ground spBO-band, which continue also at lower frequencies as the wavepacket proceeds to the right and through the avoided crossing at around $R=1.2$~a.u.. 
This is also reflected in the mixed character of R-resolved BO-population (see Fig.\ref{fig:MultiPhoton2}(a.2)), showing an increase of the ground electronic state population before reaching the avoided crossing. The part of the wavepacket already transferred to the ground state before reaching the avoided crossing of the spBO surfaces has to climb a potential hill to pass through, hence less reaches the right side. This effect, together with the weakening of the electron-nuclear nonadiabaticity from the self-polarization term distorting the BO surfaces,  yields a suppression of both the electron and proton transfer. The photon number, dominated again by the self-polarization contribution (third term in Eq.~\ref{eq:Np}), shows a corresponding increase at around $23~fs$, Fig.\ref{fig:MultiPhoton3}.

Turning now to the 30-mode case (orange)  with $\omega_{max}=0.055$~a.u., the distortion of the spBO states from the BO increases, with the avoided crossing widening slightly and shifting leftward to $R=0.5$~a.u.. The broadened one-photon bands lead to more and and earlier photon emission compared to the 10-mode case (Fig.~\ref{fig:MultiPhoton3}). The highest cavity-mode  frequencies included now are resonant with the self-polarization modified molecular resonance already at $R=-1.3$~a.u.. However, due to the change in curvature of the excited spBO surface compared to the BO surface, we observe a clear suppression of the dynamics even before the wave packet reaches this region, as evident from Fig.~\ref{fig:MultiPhoton2}.(a.1). Once the wave packet approaches $R=-1.3$~a.u. the cavity modes become resonant enabling  transitions from the $0$-photon excited spBO surface to the broadened $1$-photon ground spBO-band; the narrowing of the spBO energy differences as the wavepacket progresses past this point leads to transitions to $1$- and $(n>1)$-photon ground-spBO surfaces of the lower frequency modes. (Again, the many crossings of these surfaces become avoided crossings of the polaritonic surfaces).
In fact we see even earlier an increase in the  $R$-resolved BO ground-state population (orange in panel (a.1) in Fig.~\ref{fig:MultiPhoton2}) for the left part of the wave packet, and an increase in the photon number around $20$~fs in Fig ~\ref{fig:MultiPhoton3}. Why this happens to the left of the wavepacket rather than the right (the right is less off-resonance than the left), could be due to a stronger photon-matter coupling there from the larger molecular dipole in the left tail of the wavepacket compared to the leading edge. 
The right part of the nuclear wave packet shows a more mixed character of the $R$-resolved BO ground-state population at early times. The combined effects of increased early transitions to the electronic ground spBO state and a slightly less sharp electron-nuclear non-adiabatic region, leads to a less of the nuclear wave packet reaching the avoided crossing and a reduced electron-proton-transfer dramatically, as shown by the electronic and nuclear dipoles and the BO-occupations.

In the 50-mode case (red), the self-polarization term distorts the spBO surfaces further (Fig.~\ref{fig:MultiPhoton1}), shifting the electron-nuclear non-adiabatic region to be centered near $R = 0$~a.u.. The $1$-photon bands are wider, with $\omega_{max}=0.091$~a.u., and become resonant with the self-polarization modified molecular resonance already at $R=-3$~a.u. This leads to transitions from the initially 0-photon excited spBO surface to the $1$-photon ground spBO surfaces already at very short times.
 This is evident in the almost immediate mixed character of the $R$-resolved BO populations. The flatter slope of the excited spBO surface together with the increased population in the lower spBO surface (Fig.~\ref{fig:MultiPhoton2}.a), greatly slows the nuclear density down compared to the fewer-mode cases, and results in a full suppression of both the proton and electron transfer, as evident from panels (a-c) in Fig.~\ref{fig:MultiPhoton1}. The wave packet reflects before appreciably reaching the avoided crossing, and the change in the spatially-averaged BO-population in Fig.\ref{fig:MultiPhoton2}.(b) is caused solely by the nuclear wave packet dropping into the ground-BO state by emitting photons, and is not due to the avoided crossing at $R=0$~a.u. Considering the photon number shown in Fig.~\ref{fig:MultiPhoton2}, although 
the free photonic field component in panel (b) has a rapid initial increase and then grows throughout the time evolution as in the few-mode cases, while the linear term has a compensating decrease (panel c), the total photon number decreases after some time, due to the self-polarization contribution. As expected, this term increases with the number of modes at the initial time, but since it is proportional to the the total dipole of the system, whose transfer is suppressed, the resulting photon number tracks this behavior and is ultimately reduced compared with the fewer-mode cases.

The 70-mode case (blue) with $\omega_{max} = 0.127$~a.u. can be seen as an enhanced 50-mode case and leads to an even stronger suppression of proton-coupled electron transfer, with the same two key features that have been responsible for the cavity-modified dynamics in the 10-and higher-mode cases now having an even greater impact. First, by including the resonant frequency of the initial position of the nuclear wave packet, part of the wave packet almost immediately drops to the lower surface by emitting photons Fig.~\ref{fig:MultiPhoton3}; see also $R$-averaged BO populations at early times, Fig.~\ref{fig:MultiPhoton2}.(b) and the  mixed character developing in the $R$-resolved populations of Fig.~\ref{fig:MultiPhoton2}.(a). Second, the  deviation of the excited spBO surface is now strong enough that its gradient slopes back to the left soon after the initial nuclear wavepacket slides down from its initial position at $R = -4$~a.u., sloping back to the left, in contrast to the cavity-free excited BO surface. The overlap of the extensively broadened $1$-photon-excited- and $1$-ground-bands increases significantly creating a near-continuum of avoided crossings of polaritonic surfaces. The $0$-photon surfaces are everywhere surrounded by near-lying $n$-photon surfaces. Compared to the 50-mode case, even less density reaches the region of the avoided crossing, which is now even wider. The slope of the excited spBO-band results in even slower nuclear dynamics, with the nuclear and electronic dipole returning to their initial positions after only a small excursion away, as evident in Fig.~\ref{fig:MultiPhoton1}. Analogously to the discussion for the 50-mode case, we find a larger initial total photon number, compared to the 50-mode case, followed by a moderate increase and decrease due to the early reflection of the nuclear wave packet.

\begin{figure}[b]
\includegraphics[width=1.0\columnwidth]{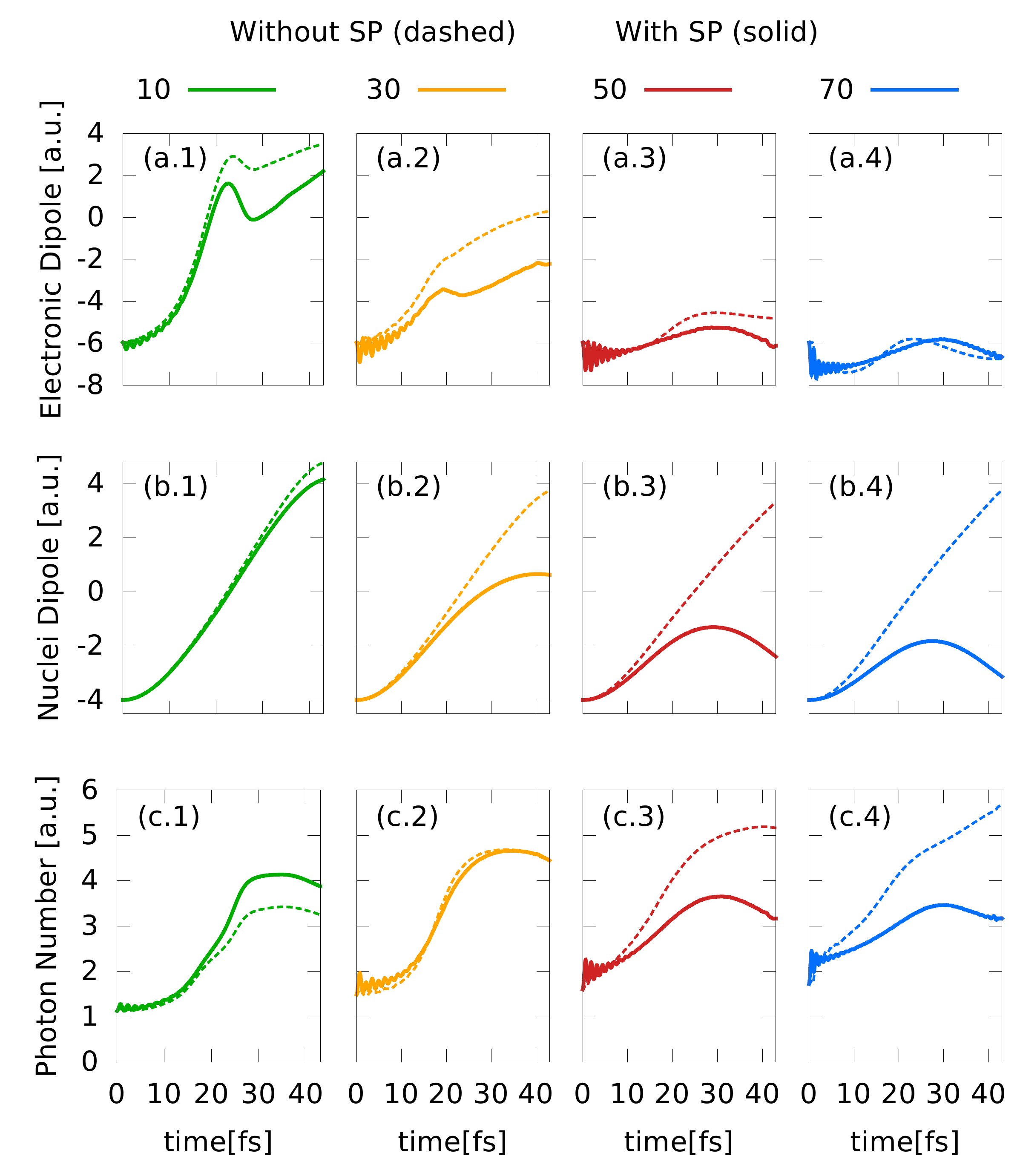}
\caption{Difference of the photon number (upper panel), nuclear dipole (middle panel) and electronic dipole (lower panel) without self-polarization term (dashed) and with self-polarization term (solid) in the same color code as Fig.\ref{fig:MultiPhoton2}.}
\label{fig:MultiPhoton_sp}
\end{figure}

Finally, to emphasize the importance of the self-polarization term on the dynamics, in Fig.~\ref{fig:MultiPhoton_sp} we compare the results of the MTE dynamics on the electronic and nuclear dipoles and photon number when this term is neglected (dashed) or included (solid) for 10, 30, 50 and 70 modes. 
 For the electronic and nuclear dipole already for the 10 mode case deviations up to $1.2$~a.u. (electronic) and $0.4$~a.u. (nuclear) are found at later times. The error in neglecting self-polarization becomes especially notable for the 50- and 70-mode cases, where including the self-polarization yields a decrease of the proton-transfer (from 50 modes to 70 modes), while not including the self-polarization yields an increase of the proton transfer. Therefore, neglecting the self-polarization term for many photon modes not only changes the quantitative results dramatically, but can also result in overall different physical effects. In fact, the nuclear and electronic wavepackets in the 70-mode case become delocalized over the entire region, so plotting simply the dipole, an averaged quantity, appears to give more agreement with the self-polarization-neglected dynamics, when in fact the wavepackets look completely different (see also supplementary material, compare movie 2. and 3.).
Turning to the total photon number, we find that the more photon modes are accounted for, the simulations without self-polarization first underestimates (10-mode), then coincides (30-modes) and then overestimates the photon number up to a factor of $2.8$ (70-modes), compared to the simulations that include self-polarization. This can be explained with the trends of the total dipole moment discussed above, since a dominant contribution to the photon number is the self-polarization term in Eq.~(\ref{eq:Np}) which depends directly on this.

\section{Discussion and Outlook}
Our results suggest that the effect of multiple cavity-modes on  the reaction dynamics can lead to dramatically different dynamics than the cavity-free case. This is true even when the cavity-modes are far from the electronic resonances encountered in the dynamics, and even more so when cavity-modes are resonant with the matter system.  In particular, for the model of cavity-induced suppression of proton-coupled electron transfer investigated here, we find  an overall increase of the suppression the more photon modes are accounted for. Two mechanisms are fundamentally responsible for the difference: First, the self-polarization term grows in significance with more modes with the effect that self-polarization-modified BO surfaces are distorted significantly away from their cavity-free shape. Polaritonic surfaces, eigenvalues of $H - T_n$, should include the explicit matter-photon coupling on top of these spBO surfaces.  Second, the $n$-photon-spBO bands become wider and increasingly overlapping, yielding a very mixed electronic character with much exchange between surfaces. These new dressed potential energy  surfaces provide a useful backdrop to analyze the dynamics, and will form a useful tool in analyzing the different surfaces put forward to study coupled photon-matter systems, for example the polaritonic surfaces, and especially the time-dependent potential energy surface arising from the exact factorization as this single surface alone provides a complete picture of the dynamics.

The MTE treatment of the photons appears to be a promising route towards treating realistic light-matter correlated systems. In particular,  this method is able to capture quantum effects such as cavity-induced suppression of proton-coupled electron transfer, yet overcomes the exponential scaling problem with the number of quantized cavity modes. However, a practical approach for realistic systems will further need an approximate treatment of the matter part. From the electronic side TDDFT would be a natural choice, while a practical treatment of nuclei calls for a classical treatment such as Ehrenfest or surface-hopping in some basis. The multiple-crossings inside the $n$-photon spBO bands suggest that simple surface-hopping treatments based on spBO surfaces should be used with much caution and that decoherence-corrections should be applied, for example those generalized from the exact factorization approach to the electron-nuclear problem~\cite{MAG15,HLM18}. Further, the MTE approach could provide a way to accurately approximate the light-matter interaction part of the QEDFT xc functional~\cite{RFPATR14,PFTAR15,FSRAR18,RTFAR18}.

Finally, we note that the present findings are general in that the increasing importance of self-polarization with more photon modes is expected to hold for the description and control of cavity-driven physical processes of molecules, nanostructures and solids embedded in cavities in general. Extensions of these findings to multi-mode cavitites suggest a new possibility of controlling and changing chemical reactions via the self-polarization without the need to explicitly change the light-matter coupling strength itself.

\begin{acknowledgments}{We would like to thank Johannes Feist for insightful discussions and the anonymous referees for very helpful comments. Financial support from the US National Science Foundation
CHE-1940333 (NM) and the Department of Energy, Office
of Basic Energy Sciences, Division of Chemical Sciences,
Geosciences and Biosciences under Award DE-SC0020044 (LL)
are gratefully acknowledged. NMH gratefully acknowledges an IMPRS fellowship.
This work was also supported by the European Research Council (ERC-2015-AdG694097), the Cluster of Excellence  (AIM), Grupos Consolidados (IT1249-19) and SFB925 "Light induced dynamics and control of correlated quantum systems. The Flatiron Institute is a division of the Simons Foundation.
}
\end{acknowledgments}

\section*{Data Availability Statement}
The data that support the findings of this study are available from the corresponding authors upon reasonable request.

\bibliography{./main_bib}

\end{document}